\documentstyle[12pt]{article}
\advance \topmargin by -\headheight
\advance \topmargin by -\headsep

\evensidemargin \oddsidemargin
\begin{document}
\topmargin -.6in

%
\def\rf#1{(\ref{eq:#1})}
\def\lab#1{\label{eq:#1}}
\def\nonu{\nonumber}
\def\br{\begin{eqnarray}}
\def\er{\end{eqnarray}}
\def\be{\begin{equation}}
\def\ee{\end{equation}}
\def\eq{\!\!\!\! &=& \!\!\!\! }
\def\foot#1{\footnotemark\footnotetext{#1}}
\def\lb{\lbrack}
\def\rb{\rbrack}
\def\llangle{\left\langle}
\def\rrangle{\right\rangle}
\def\blangle{\Bigl\langle}
\def\brangle{\Bigr\rangle}
\def\llbrack{\left\lbrack}
\def\rrbrack{\right\rbrack}
\def\lcurl{\left\{}
\def\rcurl{\right\}}
\def\({\left(}
\def\){\right)}
\newcommand{\nit}{\noindent}
\newcommand{\ct}[1]{\cite{#1}}
\newcommand{\bi}[1]{\bibitem{#1}}
\def\lskip{\vskip\baselineskip\vskip-\parskip\noindent}
\relax

\def\tr{\mathop{\rm tr}}
\def\Tr{\mathop{\rm Tr}}
\def\v{\vert}
\def\bv{\bigm\vert}
\def\Bgv{\;\Bigg\vert}
\def\bgv{\bigg\vert}
\newcommand\partder[2]{{{\partial {#1}}\over{\partial {#2}}}}
\newcommand\funcder[2]{{{\delta {#1}}\over{\delta {#2}}}}
\newcommand\Bil[2]{\Bigl\langle {#1} \Bigg\vert {#2} \Bigr\rangle}  
\newcommand\bil[2]{\left\langle {#1} \bigg\vert {#2} \right\rangle} 
\newcommand\me[2]{\left\langle {#1}\right|\left. {#2} \right\rangle} 
\newcommand\sbr[2]{\left\lbrack\,{#1}\, ,\,{#2}\,\right\rbrack}
\newcommand\pbr[2]{\{\,{#1}\, ,\,{#2}\,\}}
\newcommand\pbbr[2]{\lcurl\,{#1}\, ,\,{#2}\,\rcurl}
%
\def\a{\alpha}
\def\b{\beta}
\def\dc{{\cal D}}
\def\d{\delta}
\def\D{\Delta}
\def\eps{\epsilon}
\def\vareps{\varepsilon}
\def\g{\gamma}
\def\G{\Gamma}
\def\grad{\nabla}
\def\h{{1\over 2}}
\def\l{\lambda}
\def\L{\Lambda}
\def\m{\mu}
\def\n{\nu}
\def\o{\over}
\def\om{\omega}
\def\O{\Omega}
\def\p{\phi}
\def\P{\Phi}
\def\pa{\partial}
\def\pr{\prime}
\def\ra{\rightarrow}
\def\s{\sigma}
\def\S{\Sigma}
\def\t{\tau}
\def\th{\theta}
\def\Th{\Theta}
\def\ti{\tilde}
\def\wti{\widetilde}
\def\jc{J^C}
\def\bj{{\bar J}}
\def\sj{{\jmath}{}}
\def\bsj{{\bar \jmath}{}}
\def\bp{{\bar \p}}
\def\ca{{\cal A}}
\def\cb{{\cal B}}
\def\ce{{\cal E}}
\newcommand\sumi[1]{\sum_{#1}^{\infty}}   
%
\def\lie{{\cal G}}
\def\dlie{{\cal G}^{\ast}}
\def\elie{{\widetilde \lie}}
\def\edlie{{\elie}^{\ast}}
\def\hlie{{\cal H}}
\def\wlie{{\widetilde \lie}}
\def\f#1#2#3 {f^{#1#2}_{#3}}
\def\winf{{\sf w_\infty}}
\def\win1{{\sf w_{1+\infty}}}
\def\hwinf{{\sf {\hat w}_{\infty}}}
\def\Winf{{\sf W_\infty}}
\def\Win1{{\sf W_{1+\infty}}}
\def\hWinf{{\sf {\hat W}_{\infty}}}
%
%
\newcommand\PRL[3]{{\sl Phys. Rev. Lett.} {\bf#1} (#2) #3}
\newcommand\NPB[3]{{\sl Nucl. Phys.} {\bf B#1} (#2) #3}
\newcommand\NPBFS[4]{{\sl Nucl. Phys.} {\bf B#2} [FS#1] (#3) #4}
\newcommand\CMP[3]{{\sl Commun. Math. Phys.} {\bf #1} (#2) #3}
\newcommand\PRD[3]{{\sl Phys. Rev.} {\bf D#1} (#2) #3}
\newcommand\PLA[3]{{\sl Phys. Lett.} {\bf #1A} (#2) #3}
\newcommand\PLB[3]{{\sl Phys. Lett.} {\bf #1B} (#2) #3}
\newcommand\JMP[3]{{\sl J. Math. Phys.} {\bf #1} (#2) #3}
\newcommand\PTP[3]{{\sl Prog. Theor. Phys.} {\bf #1} (#2) #3}
\newcommand\SPTP[3]{{\sl Suppl. Prog. Theor. Phys.} {\bf #1} (#2) #3}
\newcommand\AoP[3]{{\sl Ann. of Phys.} {\bf #1} (#2) #3}
\newcommand\PNAS[3]{{\sl Proc. Natl. Acad. Sci. USA} {\bf #1} (#2) #3}
\newcommand\RMP[3]{{\sl Rev. Mod. Phys.} {\bf #1} (#2) #3}
\newcommand\PR[3]{{\sl Phys. Reports} {\bf #1} (#2) #3}
\newcommand\AoM[3]{{\sl Ann. of Math.} {\bf #1} (#2) #3}
\newcommand\UMN[3]{{\sl Usp. Mat. Nauk} {\bf #1} (#2) #3}
\newcommand\FAP[3]{{\sl Funkt. Anal. Prilozheniya} {\bf #1} (#2) #3}
\newcommand\FAaIA[3]{{\sl Functional Analysis and Its Application} {\bf #1}
(#2) #3}
\newcommand\BAMS[3]{{\sl Bull. Am. Math. Soc.} {\bf #1} (#2) #3}
\newcommand\TAMS[3]{{\sl Trans. Am. Math. Soc.} {\bf #1} (#2) #3}
\newcommand\InvM[3]{{\sl Invent. Math.} {\bf #1} (#2) #3}
\newcommand\LMP[3]{{\sl Letters in Math. Phys.} {\bf #1} (#2) #3}
\newcommand\IJMPA[3]{{\sl Int. J. Mod. Phys.} {\bf A#1} (#2) #3}
\newcommand\AdM[3]{{\sl Advances in Math.} {\bf #1} (#2) #3}
\newcommand\RMaP[3]{{\sl Reports on Math. Phys.} {\bf #1} (#2) #3}
\newcommand\IJM[3]{{\sl Ill. J. Math.} {\bf #1} (#2) #3}
\newcommand\APP[3]{{\sl Acta Phys. Polon.} {\bf #1} (#2) #3}
\newcommand\TMP[3]{{\sl Theor. Mat. Phys.} {\bf #1} (#2) #3}
\newcommand\JPA[3]{{\sl J. Physics} {\bf A#1} (#2) #3}
\newcommand\JSM[3]{{\sl J. Soviet Math.} {\bf #1} (#2) #3}
\newcommand\MPLA[3]{{\sl Mod. Phys. Lett.} {\bf A#1} (#2) #3}
\newcommand\JETP[3]{{\sl Sov. Phys. JETP} {\bf #1} (#2) #3}
\newcommand\JETPL[3]{{\sl  Sov. Phys. JETP Lett.} {\bf #1} (#2) #3}
\newcommand\PHSA[3]{{\sl Physica} {\bf A#1} (#2) #3}
\newcommand\PHSD[3]{{\sl Physica} {\bf D#1} (#2) #3}
\def\vo{V^{(0)}}
\def\vp{V^{(p)}}
\def\vu{V^{(1)}}

\begin{titlepage}
\vspace*{-1cm}
\noindent
December, 1993 \hfill{IFT-P/75/93}\\
\phantom{bla}
\hfill{UICHEP-TH/93-17}\\
\phantom{bla}
\hfill{hep-th/9312144}
\\
\vskip .3in

\begin{center}

{\large\bf On Discrete Symmetries of the  }
\end{center}
\begin{center}
{\large\bf Multi-boson KP Hierarchies}
\end{center}
\normalsize
\vskip .4in

\begin{center}
{ H. Aratyn\footnotemark
\footnotetext{Work supported in part by U.S. Department of Energy,
contract DE-FG02-84ER40173 and by NSF, grant no. INT-9015799}}

\par \vskip .1in \noindent
Department of Physics \\
University of Illinois at Chicago\\
845 W. Taylor St.\\
Chicago, Illinois 60607-7059\\
\par \vskip .3in

\end{center}

\begin{center}
{L.A. Ferreira\footnotemark
\footnotetext{Work supported in part by CNPq}} and A.H. Zimerman$^{\,2}$

\par \vskip .1in \noindent
Instituto de F\'{\i}sica Te\'{o}rica-UNESP\\
Rua Pamplona 145\\
01405-900 S\~{a}o Paulo, Brazil
\par \vskip .3in

\end{center}

\begin{center}
{\large {\bf ABSTRACT}}\\
\end{center}
\par \vskip .3in \noindent

We show that the multi-boson KP hierarchies possess a class of
discrete symmetries linking them to the discrete Toda systems. These discrete
symmetries are generated by the similarity transformation of the corresponding
Lax operator. This establishes a canonical nature of the
discrete transformations.
The spectral equation, which defines both the lattice system and
the corresponding Lax operator, plays a key role in determining pertinent
symmetry structure.

We also introduce a concept of the square-root lattice leading to a family
of new pseudo-differential operators with covariance under
additional B\"{a}cklund transformations.

\end{titlepage}
{\large {\bf 1. Introduction}}
\lskip

Two-boson Lax operator emerged naturally as an operator defining
closed subspace of the space dual to the differential operators \ct{R82}
and was studied first (mostly in its quasi-classical limit)
in connection with an hierarchy of Benney equations \ct{LM79,BAK85}.
More recently the two-boson hierarchy
appeared in connection with ${\sf W}$-infinity algebras, CAT model and
one-matrix models \ct{YW9111,2boson,BX9204,BX9209}.
A study of Adler-Kostant-Symes (AKS) approach to various integrable models
of the KP type revealed existence of the four-boson KP hierarchy which via
symplectic Miura-like gauge transformation formed another representation of
$\Win1$ algebra \ct{ANPV}.
Both two- and four-boson representations
were consistent reductions of the full KP hierarchy understood as coadjoint
orbit system within AKS scheme \ct{ANPV}.
Investigations of the multi-matrix models and corresponding Toda
hierarchy led to a class of the pseudodifferental operators
\ct{BX9212,BX9305}, which although formally generalized
two-and four boson systems contained an arbitrary even number of fields,
and did not arise as a finite orbit within the AKS scheme.
Their status as a consistent hamiltonian reduction of the full KP hierarchy
was established in \ct{ANP93}, where the structure of their first Poisson
bracket was found.

Recently, it was pointed out that the continuous two-boson KP hierarchy
possesses a class of discrete canonical symmetries \ct{discrete}.
The symmetry transformations were found to be associated with the
translational shift on the corresponding Toda lattice \ct{discrete,LSY93}.
Also, it was shown how due to the presence of this symmetry one
can construct the site functions of the lattice model as gauge copies of the
continuous KP model.
In this letter the canonical nature of the discrete symmetry for a wide class
of multi-boson KP hierarchies is being
explained as we show that the discrete symmetries are generated by the
similarity transformations of the Lax operator.

We discuss the link between discrete and continuous model by means of
the linear spectral equation.
The symmetries of the integrable model can be associated to transformations
which keep the eigenvalue of the spectral equations invariant.
In cases where the spectral equation allows the finer lattice structure,
which we call a ``square-lattice", we find additional symmetries; so-called
generalized B\"{a}cklund transformations.
These transformations operate on a new class of multi-boson
pseudo-differential operators, which generalize Volterra lattice
known for the two-boson case.
\lskip
{\large {\bf 2. From Continuum KP Multi-boson systems to Generalized Toda
Chains via Similarity Transformations}}
\lskip
Here we construct the lattice
system with lattice shifting generated by the similarity transformation
acting on the continuous Lax of the type:
\be
L^{(N)} \equiv \pa +
\sum_{k=1}^{N} a_k {1 \o \pa - S_k} \cdots {1 \o \pa - S_2}{1 \o \pa - S_1}
\lab{ln}
\ee
for any finite integer $N \geq 1$. Introduce
$S_N^0 \equiv S_N+ \pa \ln a_N $ and consider the following transformation
of \rf{ln}:
\be
{\wti L}^{(N)} = \( \pa - S^0_N \) \; L^{(N)} \; \( \pa - S^0_N \)^{-1}
\lab{transln}
\ee
Using obvious identities $(\pa - \p ) (\pa - \psi)^{-1} = (\psi - \p )
(\pa - \psi)^{-1} + 1$ and $\p  (\pa - \psi)^{-1} =
(\pa - \psi-\pa \ln \p )^{-1} \p$ we find:
\br
{\wti L}^{(N)} \eq \pa +  (\pa S_N^0) {1 \o \pa - S_N^0} + \sum_{k=1}^{N}
a_k {1 \o \pa - S_{k-1}}\cdots {1 \o \pa - S_1}{1 \o \pa - S_N^0} \nonu\\
&+& \sum_{k=1}^{N-1} \lb (\pa a_k) + a_k \( S_k - S_N^0 \) \rb
{1 \o \pa - S_{k}} \cdots {1 \o \pa - S_1} {1 \o \pa - S_N^0}
\lab{lntrans}
\er
which again has a form of \rf{ln}:
\be
{\wti L}^{(N)} = \pa +
\sum_{k=1}^{N} {\wti a_k} {1 \o \pa - {\wti S}_{k}} \cdots
{1 \o \pa - {\wti S}_1}  \lab{tiln}
\ee
with
\br
{\wti a}_1 \eq a_1 + (\pa S_N^0 ) \lab{tiaone}\\
{\wti a}_{l+1} \eq  a_{l+1} + \pa a_l - a_l \( S_N - S_l + \pa \ln a_N \)
\qquad l = 1,2,\ldots,N-1          \lab{tial}\\
{\wti S}_1  \eq S_N^0 = S_N + \pa \ln a_N \lab{tisone}\\
{\wti S}_k \eq S_{k-1} \qquad k = 2,\ldots,N \lab{tisk}
\er
we will see below connection to the multi-boson Toda type lattice, where
transformation \rf{tiaone}-\rf{tisk} will find realization as a shift of
the Toda lattice.

In reference \ct{ANP93} it was shown that the Lax operators
\rf{ln} are consistent Poisson reductions of the general KP Lax operator
and have their first bracket structure given by:
\be
\lcurl \llangle L^{(N)} \bv X \rrangle \, ,\,
\llangle L^{(N)} \bv Y \rrangle \rcurl_1 =
\llangle L^{(N)} \bv \left\lb X,\, Y
\right\rb \rrangle    \lab{3-6}
\ee
Since the transformed Lax \rf{tiln} is of the same form as \rf{ln} it
should obey \rf{3-6} as well.
This establishes symplectic character of the similarity transformation
\rf{transln} with respect to the first Poisson bracket structure.
Moreover both systems are consistent AKS systems as
reductions of the full KP hierarchy. Accordingly their common set of
Hamiltonians are in involution, which ensures integrability of both systems.
Furthermore invariance of the higher bracket
follows by application of the Lenard relations.

Note, that the transformation \rf{transln} can also be written as
$G^{-1} \, {\wti L}^{(N)}\, G = \pa \; G^{-1} L^{(N)} G \; \pa^{-1}$,
with $ G \equiv \exp \(\int S_N^0 \)$.
\lskip
{\large {\bf 3. Symmetries of the Linear Spectral System.}}
\lskip
The idea here is to look for symmetries of the spectral system defined by two
equations:
\br
\pa \Psi_n \eq \Psi_{n+1} + a_0 (n) \Psi_n     \lab{speca}\\
\l \Psi_n \eq \Psi_{n+1}  + a_0 (n) \Psi_n
+ \sum_{k=1}^N a_k (n) \Psi_{n-k} \; = \; L_n^{(N)} \, \Psi_{n} \qquad
\; \; \forall n
\lab{specb}
\er
where
\be
L_n^{(N)} = \pa + \sum_{k=1}^N\, a_k (n)\, {1 \o \pa - a_0 (n-k)}
\ldots {1 \o \pa - a_0 (n-1)}
\lab{lnn}
\ee
To obtain \rf{lnn} we used an inverted version of \rf{speca}:
\be
\Psi_n = {1 \o \pa - a_0 (n)} \Psi_{n+1} \to
\Psi_{n-j} = {1 \o \pa - a_0 (n-j)} {1 \o \pa - a_0 (n-j+1)}
\ldots {1 \o \pa - a_0 (n-1)} \Psi_n
\lab{psinminusj}
\ee
{\bf Symmetries:}~
We now define as \underbar{symmetry} of the system \rf{speca}-\rf{specb}
any transformation which leaves the eigenvalue $\l$ invariant.
The reason is that then $\Tr (L^n)$ would be unchanged too (for the
appropriate definition of trace).

One finds two classes of symmetry transformations.

{\bf Discrete Symmetry:}~ Since from \rf{specb} we also have:
$\l \Psi_{n+1} = L_{n+1}  \Psi_{n+1} $, thus the transformation
$n \to n+1$ does not change the eigenvalue $\l$.
Combining with \rf{speca} and \rf{specb} we find
\br
L_{n+1}^{(N)} \Psi_{n+1} &=& \l \( \pa - a_0 (n) \) \Psi_n =
 \( \pa - a_0 (n) \) L_n^{(N)} \Psi_n \nonu\\
&=& \( \pa - a_0 (n) \) L_n^{(N)} \( \pa - a_0 (n) \)^{-1} \Psi_{n+1}
\lab{lntoln1}
\er
So, the Lax at different sites are related by a similarity transformation:
\be
L_{n+1}^{(N)} = \( \pa - a_0 (n) \)\; L_n^{(N)}\; \( \pa - a_0 (n) \)^{-1}
\lab{transf3}
\ee
Since the Lax operators at different sites are connected via
similarity transformation, the Hamiltonians calculated by Adler trace are
the same!
Moreover the consistency of \rf{speca}- \rf{specb} requires
$a_0 (n) = a_0 (n-N) + \pa \ln a_N (n)$, which establishes a link with
result of the previous section.

{\bf Phase Symmetry:}~ Consider $L_{n}^{\pr} = \exp ( \Phi) L_{n}
\exp ( -\Phi)$. Then
\be
\l \Psi_{n} = L_{n}  \Psi_{n} \; \to \;
\l \( e^{\Phi} \Psi_{n}\)  = L_{n}^{\pr} \( e^{\Phi} \Psi_{n}\)
\lab{gaugesym}
\ee
determines another symmetry of the system given by \rf{speca}- \rf{specb}
(see \ct{ANPV} for discussion of its symplectic character).

{\bf Example. Four-boson system.} We now focus on four-boson Lax operator
as described by Bonora-Xiong (BX) in \ct{BX9212}:
\be
L_n^{(2)} \equiv \pa + a_1 (n) {1 \o \pa - a_0 (n-1)} +
a_2 (n) {1 \o \pa - a_0 (n-2)}{1 \o \pa - a_0 (n-1)}
\lab{lj2}
\ee
where $\pa = \pa / \pa t_{1,1}$. A partial derivative with respect to
another time direction $t_{q,1}$ is denoted by ${\ti \pa} \equiv
 \pa / \pa t_{q,1}$.
According to BX $L_n$ is obtained from $t_{1,1}$ flow
of $\Psi_n$ together with spectral equations \rf{psia} and
\rf{psic}:
\br
\pa \Psi_n \eq \Psi_{n+1} + a_0 (n) \Psi_n     \lab{psia}\\
{\ti \pa} \Psi_n \eq - R_n \Psi_{n-1}                \lab{psib}\\
\l \Psi_n \eq \Psi_{n+1}  + a_0 (n) \Psi_n
+ a_1 (n) \Psi_{n-1}   + a_2 (n) \Psi_{n-2}   \lab{psic}
\er
Let us now discuss consequences of \rf{psia}-\rf{psic} being a consistent set
of equations with $\pa \l =0 $.
Consistency of \rf{psia} and \rf{psic} yields the following evolution
equations in $t_{1,1}$:
\br
\pa a_0 (n) \eq a_1 (n+1) - a_1 (n)  \lab{t1eqa} \\
\pa a_1 (n) \eq a_2 (n+1) - a_2 (n)  + a_1 (n) \( a_0 (n) - a_0 (n-1) \)
           \lab{t1eqb}\\
\pa a_2 (n) \eq a_2 (n)  \( a_0 (n) - a_0 (n-2) \)       \lab{t1eqc}
\er
while for the flows in $t_{q,1}$ we get from requiring consistency of
\rf{psib}-\rf{psic}:
\br
{\ti \pa} a_0 (n) \eq R_{n+1} - R_{n} \quad;\quad
{\ti \pa} a_1 (n) = R_n \( a_0 (n) - a_0 (n-1) \) \lab{tq1eqa} \\
{\ti \pa} a_2 (n) \eq a_1 (n) R_{n-1} - a_1 (n-1) R_n
\quad;\quad R_n a_2 (n-1) = a_2 (n) R_{n-2} \lab{tq1eqc}
\er
(the last equation in \rf{tq1eqc} is a condition and not evolution equation
because $a_3$ is absent in this system).
On top of this there is an evolution equation for $R_n$, which follows
from consistency of \rf{psia}-\rf{psib}:
\be
\pa R_n =  R_n  \( a_0 (n) - a_0 (n-1) \)       \lab{rjevo}
\ee
with this equation $R_n$ satisfies the 2-dimensional Toda lattice
equation as shown in \ct{BX9212}.
Solution to the last equation in \rf{tq1eqc} is $a_2 (n) = R_n R_{n-1}$
which not only satisfies this equation but is also consistent with remaining
equation in \rf{t1eqc} and also \rf{rjevo}.

To see better what is happening let us for the moment go back to
the simpler case of the two-boson system. Since there $a_2 =0$ eq.\rf{tq1eqc}
forces $R_n = a_1 (n)$ and with this we are left with two modes $a_0$ and
$a_1$. Furthermore two set of evolution equations
become equal and we effectively have one flow parameter $t_{1,1}=t_{q,1}$.
The remaining two evolution equations define the shift which turns out to be a
canonical symmetry of the two-boson KP system. More explicit we have a
two-boson Lax:
\be
L_n^{(1)} \equiv \pa + a_1 (n) {1 \o \pa - a_0 (n-1)}
\lab{lj1}
\ee
and the symmetry operations:
\br
a_0 (n) \eq a_0 (n-1) + \pa \ln a_1 (n)             \lab{2ba}\\
a_1 (n+1) \eq a_1 (n) + \pa a_0 (n)=a_1 (n) + \pa a_0 (n-1) +
\pa^2 \ln a_1 (n)             \lab{2bb}
\er
which become a $G^{-1}$ symmetry transformation
\be
G^{-1} (J) = J - \( \ln \bj \)^{\pr}
\quad \quad\quad\quad\quad
\mbox{\rm and}    \qquad \quad \quad G^{-1} (\bj) =
\bj + \( \ln \bj  \)^{\pr \pr} - J^{\pr}
\lab{jtransf2}
\ee
valid for 2-boson systems as found in \ct{discrete} when we identify
$J =- a_0 (n-1), \bj=a_1 (n)$.

We deal with system defined by $a_2,a_1, a_0$ modes
treating $a_0 (n-1) \equiv S_1 (n) $ and $a_0 (n-2) \equiv S_2 (n) $ in
$L_n^{(2)}$ as different modes and hence we find:
\br
a_1 (n+1) \eq a_1 (n) + \pa S_1 (n+1)  \lab{4teqa} \\
a_2 (n +1) \eq  a_2 (n)  + \pa a_1 (n) - a_1 (n) \( S_1 (n+1) - S_1 (n) \)
           \lab{4teqb}\\
S_1 (n+1) \eq S_2 (n) + \pa \ln a_2 (n) \lab{4teqc} \\
S_2 (n+1) \eq S_1 (n) \lab{4teqd}
\er
We now associate to $L_n^{(2)}$ a continuous Lax $L^{(2)}$ :
\be
L^{(2)} \equiv \pa + a_1 {1 \o \pa - S_1} +
a_2 {1 \o \pa - S_2}{1 \o \pa - S_1}
\lab{l2}
\ee
and associate the discrete shifts from \rf{4teqa}-\rf{4teqd} to
the following symmetry operations of continuous variables:
\br
a_1^{\pr} \eq a_1 + \pa \( S_2 + \pa \ln a_2 \)
\quad;\quad
a_2^{\pr} = a_2 + \pa a_1 - a_1 \( S_2 - S_1 + \pa \ln a_2 \)
\lab{disca}\\
S_1^{\pr}  \eq S_2 + \pa \ln a_2  \quad;\quad
S_2^{\pr} = S_1 \lab{discd}
\er
which is a special case of \rf{tiaone}-\rf{tisk}.
An explicit calculation verified that both first and second bracket
structures are invariant in agreement with our general proposition
proved in sect. 2.
Accordingly, due to theorem from \ct{discrete} this establishes
\rf{disca}-\rf{discd} as a canonical mapping, leaving invariant
in form \underbar{all} bracket structures.

{\bf General case:~}For N-bose model we have correspondingly $
a_N (n) = R_n R_{n-1}\cdots R_{n-N+1}$ with evolution equations:
\br
\pa a_0 (n) \eq a_1 (n+1) - a_1 (n)  \lab{nt1eqa} \\
\pa a_l (n) \eq a_{l+1} (n+1) - a_{l+1} (n)  + a_l (n)
\( a_0 (n) - a_0 (n-l) \) \quad \; l=0,1,2,\ldots,N-1 \lab{nt1eqb}\\
\pa a_N (n) \eq a_N (n)  \( a_0 (n) - a_0 (n-N) \)       \lab{nt1eqc}
\er
and $a_0 (n-N) \equiv S_N (n),\ldots, a_0 (n-2)\equiv S_2 (n),
a_0 (n-1)\equiv S_1 (n)$ enter the system together with
$a_1 (n) ,a_2 (n), \ldots, a_N (n)$ as independent modes.
These equations translate to the following discrete symmetry pattern:
\br
a_1^{\pr} \eq a_1 + \pa \( S_N + \pa \ln a_N \) \lab{gdisca}\\
a_{l+1}^{\pr} \eq  a_{l+1} + \pa a_l - a_l \( S_N - S_l + \pa \ln a_N \)
\qquad l = 1,2,\ldots,N-1          \lab{gdiscb}\\
S_1^{\pr}  \eq S_N + \pa \ln a_N \qquad;\quad\;
S_k^{\pr} = S_{k-1} \qquad k = 2,\ldots,N
\lab{gdiscc}
\er
which agree with \rf{tiaone}-\rf{tisk} for the variables appearing
in the continuous Lax operator \rf{ln}.
\lskip
{\large {\bf 4. Multi-Boson Lattice Systems and the B\"{a}cklund
Transformations.}}
\lskip
Recall first the canonical Toda Hamiltonian:
\be
H_T = \sum_{n} \( \h P_n^2 + e^{Q_{n} - Q_{n-1} } \) =
\sum_{n} \( \h S_n^2 + R_n \)
\lab{hamtoda}
\ee
where we have introduced $ S_n = P_n= \pa_t Q_n$ and $R_n
= \exp (Q_{n} - Q_{n-1})$ on the right hand side of \rf{hamtoda}.
With the canonical relation $\pbr{P_n}{Q_m} = \d_{n,m}$ the Toda equations
of motion are in terms of $S_n , R_n$:
\be
\pa_t \,S_n =  R_{n +1} - R_n  \qquad;\qquad
\pa_t\, R_n = R_n \( S_{n} - S_{n-1} \) \lab{sneq}
\ee
These equations can also be obtained as a consistency of the spectral system
(f.i. \ct{KM92})
\br
\pa \Psi_n \eq \Psi_{n+1} + S_n \Psi_n     \lab{spectoda}\\
\l \Psi_n \eq \Psi_{n+1}  + S_n \Psi_n
+ R_n \Psi_{n-1}      \lab{spectodb}
\er
which is a special two-boson case of the system \rf{speca}-\rf{specb}.
We now introduce an useful concept of a square-lattice.
Let us introduce in addition to $\Psi_n $ ``living" on integer lattice
an extra wave function ${\wti \Psi}_{n+\h} $ defined on half-integer
lattice. We can now define a ``square-root" of \rf{spectodb} by
\be
\l^{1/2}\; {\wti \Psi}_{n+\h} = \Psi_{n+1} + A_{n+1} \Psi_n
\qquad;\qquad
\l^{1/2} \; \Psi_n = {\wti \Psi}_{n+\h} + B_n {\wti \Psi}_{n-\h}
\lab{sqra}
\ee
One notices an obvious symmetry, which we here call a $g$-symmetry,
connecting both equations of \rf{sqra}:
\be
\Psi_n \to {\wti \Psi}_{n+\h} \qquad {\wti \Psi}_{n+\h}
\to \Psi_{n +1} \qquad B_n \to A_{n+1} \qquad A_{n} \to B_{n}
\lab{psitotipsi}
\ee
Introducing a new lattice system through $V_{2n} = B_n ,V_{2n-1} = A_{n}$
and $\Phi_{2n} = \Psi_n,\Phi_{2n-1} = {\wti \Psi}_{n-\h}$
we can rewrite \rf{sqra} as
\be
\l^{1/2} \; \Phi_n =  \Phi_{n+1} + V_n  \Phi_{n-1}
\lab{gensqrab}
\ee
By adding the time evolution equation:
\be
\pa \; \Phi_n =  \Phi_{n+2} + \( V_n  + V_{n+1} \) \Phi_{n}
\lab{paphin}
\ee
we find that a consistency condition for the last two equations takes a
form of the Volterra equation:
\be
\pa V_n = V_n (V_{n+1} - V_{n-1} )   \lab{vvolta}
\ee
and notice that the $g$-symmetry becomes $g \( V_{n} \)= V_{n+1}$ and
$g \( \Phi_{n}\) = \Phi_{n+1}$.
Combining both equations of \rf{sqra} we get:
\br
\l \Psi_n \eq \Psi_{n+1} + \( A_{n+1} + B_n \) \Psi_n
+ A_{n}B_n \Psi_{n-1}\lab{sqr2a}\\
\l {\ti \Psi}_{n+\h} \eq {\ti \Psi}_{n+3/2} +
\( A_{n+1} + B_{n+1} \) {\ti \Psi}_{n+\h} + A_{n+1}B_n {\ti \Psi}_{n-\h}
\lab{sqr2b}
\er
First, comparing \rf{sqr2a} and \rf{sqr2b} with \rf{spectodb} and taking into
account \rf{psitotipsi} we arrive at:
\be
S_n = B_n + A_{n+1} \quad \quad R_n = A_n B_n
\quad\; ; \quad\;
{\wti S}_{n} = B_{n+1} + A_{n +1} \quad \quad {\wti R}_{n} = A_{n +1} B_n
\lab{snab}
\ee
where the quantities $ {\wti S}_{n} = {\wti P}_n = \pa_t {\wti Q}_n $ and
${\wti R}_{n} = \exp ({\wti Q}_{n} - {\wti Q}_{n-1})$ define another Toda
system.
Toda equations of motion \rf{sneq} agree now with
the Volterra equations written in components:
\be
\pa A_n = A_n (B_n - B_{n-1} )  \qquad;\qquad
\pa B_n = B_n (A_{n+1} - A_{n} )  \lab{volta}
\ee
We will now determine Volterra variables in terms of canonical variables of
the two-boson Toda systems.
Consider
\be
S_n - {\wti S}_{n-1} = \pa_t  Q_n - \pa_t {\wti Q}_{n -1} =
A_{n +1} - A_{n}= \pa_t \ln B_n             \lab{sntisn}
\ee
from which we find $
B_n = -  e^{Q_n - {\wti Q}_{n -1}}/A$ with $A$ being a constant.
Similarly from $R_n - {\wti R}_{n-1} = A_n ( B_n - B_{n-1})$
we get $A_n = - A\; \exp \({\wti Q}_{n -1} - Q_{n-1}\)$.
Now we can comment on connection to a B\"{a}cklund transformation
\ct{To76,Wa76}.
Define a generating function
$W (Q,{\wti Q}) = \sum_n \( A_n + B_n + {\rm const.}
( {\wti Q}_n - Q_n ) \)$.
The corresponding canonical transformation is:
$P_n = \partder{W}{Q_n}$ and ${\wti P}_n =- \partder{W}{{\wti Q}_n}$
and transforms the Hamiltonian \rf{hamtoda} to
${\wti H}_T \({\wti Q}, {\wti P}\) =  H_T \(Q, P\)
+ {\rm constant}$.
Therefore as a part of equations of motion we have
${\wti P}_n = \pa_t {\wti Q}_n$ and the above canonical transformations yield
$$
\pa_t  Q_n - \pa_t {\wti Q}_{n -1} = A \(e^{{\wti Q}_{n} - Q_{n}}
- e^{{\wti Q}_{n -1} - Q_{n-1}}\)
\;\;;\;\;
 \pa_t  {\wti Q}_{n -1} - \pa_t  Q_{n -1} = {e^{Q_{n-1} - {\wti Q}_{n -2}}
 -e^{Q_{n} - {\wti Q}_{n -1}}  \o A}
$$
These equations are identical to \rf{volta}. On the other hand they
transform a solution $Q(t)$ to another solution ${\wti Q} (t)$.
This transformation is known as a B\"{a}cklund transformation and we find that
it is equivalent to Volterra equations \rf{volta} and have
therefore an origin in existence of the square-lattice.

The discrete $g$-symmetry for the system in \rf{volta} is given by
\rf{psitotipsi} and corresponds to $ S_n , R_n \to {\wti S}_{n}, {\wti R}_{n}$
as seen from \rf{snab}.
Two successive B\"{a}cklund transformations (i.e.  $g^2$) amount to
$ S_n , R_n \to S_{n+1}, R_{n+1}$, i.e. shift by the original lattice
spacing of the Toda lattice.

The challenge is now to find a square-lattice (B\"{a}cklund transformation)
for the multi-boson KP system. This will be done below.

{\bf $g$-symmetry as a similarity transformation}~ Let us go back to the time
evolution equation \rf{paphin} rewriting it in components as:
\be
{\wti \Psi}_{n+\h} = \( \pa - B_n- A_{n} \){\wti \Psi}_{n-\h}
\qquad;\qquad
\Psi_{n+1} = \( \pa - B_n- A_{n+1} \) \Psi_{n}
\lab{evolution}
\ee
With these equations we can now rewrite the ``square-root" spectral
equations \rf{sqra} as:
\be
\l^{1/2}\, {\wti \Psi}_{n+\h} = \( \pa - B_n \) \Psi_n
\qquad;\qquad
\l^{1/2} \, \Psi_n = \( \pa - A_n \) {\wti \Psi}_{n-\h}
\lab{newsqr}
\ee
If we now write \rf{newsqr} for $n+1$ and use
\rf{evolution} to go back to $n$ we find the following
relations
\br
\( \pa - B_{n+1} \) \eq \( \pa - B_{n+1} - A_{n+1} \) \( \pa - B_n \)
\( \pa - B_n- A_{n+1} \)^{-1} \lab{bnbnplus} \\
\( \pa - A_{n+1} \) \eq \( \pa - B_{n} - A_{n+1} \)\( \pa - A_n \)
\( \pa - B_{n} - A_{n} \)
\lab{ananplus}
\er
Combining two equations in \rf{newsqr} we find
\be
\l \; \Psi_n  = \( \pa - A_n \) \( \pa - B_{n-1} \) \Psi_{n-1}
= \( \pa - A_n \) \( \pa - B_{n-1} \) \( \pa - B_{n-1} - A_{n} \)^{-1}
\Psi_{n} \equiv L^{(2)}_n \Psi_{n}
\lab{voll2}
\ee
using \rf{bnbnplus} and an identity $(\pa - \psi)^{-1} (\pa - \p )=
(\pa - \psi)^{-1} (\psi - \p )+ 1$ we find:
\be
L^{(2)}_n = \( \pa - A_n \) \( \pa - B_{n} - A_{n} \)^{-1}
\( \pa - B_{n} \) = \pa + B_n \( \pa - B_{n} - A_{n} \)^{-1} A_n
\lab{wuyu}
\ee
which for $B_n = \bsj\,$ and $A_n = \sj\,$ agree with the Lax operator
of the two-boson hierarchy from \ct{YW9111}.
$L^{(2)}_n$ is equal to the Lax we have seen in the Toda section up
to a simple phase transformation \ct{AFGMZ}.
Similarly we find repeating analogous steps
\be
\l \; {\wti \Psi}_{n+\h}  = \( \pa - B_n \) \( \pa - A_{n} \)
\( \pa - B_{n} - A_{n} \)^{-1} {\wti \Psi}_{n+\h}
\equiv {\wti L}^{(2)}_n {\wti \Psi}_{n+\h}
\lab{tivoll2}
\ee
with ${\wti L}^{(2)}_n $ being equal (after use of \rf{ananplus}) to:
\be
{\wti L}^{(2)}_n = \( \pa - B_n \) \( \pa - B_{n} - A_{n+1} \)^{-1}
\( \pa - A_{n+1} \) = \pa + A_{n+1}  \( \pa - B_{n} - A_{n+1} \)^{-1}
B_n
\lab{tiwuyu}
\ee
Clearly $ g \( L^{(2)}_n \) = {\wti L}^{(2)}_n $.
To express this relation in form of the similarity relation let us go back
to $\l \Psi_n=  L^{(2)}_n \Psi_{n} $ and rewrite it using \rf{newsqr}
as
\be
\l \l^{\h} \( \pa - B_n \)^{-1} {\wti \Psi}_{n+\h} =
L^{(2)}_n \l^{\h} \( \pa - B_n \)^{-1} {\wti \Psi}_{n+\h}
\lab{step}
\ee
from which we find:
\be
{\wti L}^{(2)}_n = \( \pa - B_n \) L^{(2)}_n  \( \pa - B_n \)^{-1}
\lab{lntotiln}
\ee
which is easy to verify directly using \rf{bnbnplus} and
\rf{ananplus}.
Similarly we find
\be
L^{(2)}_{n+1}  = \( \pa - A_{n+1} \) {\wti L}^{(2)}_n  \( \pa - A_{n+1}\)^{-1}
\lab{tilntolnplus}
\ee
In the compact notation we have:
\br
L^{(2)}_n \eq \( \pa - V_{n-1} \) \( \pa - V_{n-1} - V_{n} \)^{-1}
\( \pa - V_{n} \) \lab{vln}\\
L^{(2)}_{n+1}  \eq \( \pa - V_{n} \) L^{(2)}_n  \( \pa - V_{n}\)^{-1}
\lab{vlntovlnplus}
\er
{\bf Multi-boson Volterra System.}~
We now define a general multi-boson Volterra system by the spectral
equations:
\br
\pa \; \Phi_n \eq \Phi_{n+2} + \( V^{(0)}_n  + V^{(0)}_{n+1} \) \Phi_{n}
\lab{npaphin}\\
\l^{1/2} \; \Phi_n \eq  \Phi_{n+1} + \sum_{p=0}^N \vp_{n-2p} \Phi_{n-2p-1}
\lab{genspec}
\er
with the component identification $\vp_{2n} = B^{(p)}_n$, $\vp_{2n-1} =
A^{(p)}_n$ as well as a $g$-symmetry: $g \( \vp_{n}\) = \vp_{n+1}$
(in components $B^{(p)}_n \to A^{(p)}_{n+1},
A^{(p)}_{n} \to B^{(p)}_{n}$).

Rewriting \rf{genspec} in components and taking their combination we arrive
at the corresponding Toda lattice equation for $2N+2$ Toda site functions
\be
\l \Psi_n = \Psi_{n+1}  + \sum_{k=0}^{2N+1} \( B^{(k)}_{n-k} + A^{(k)}_{n-k+1}
+ \sum_{k=m+p+1} B^{(p)}_{n-p} A^{(m)}_{n-p-m} \) \Psi_{n-k} \lab{genvotoda}
\ee
Note the number of the Toda site functions $a_0,a_1, \ldots,a_{2N+1}$ defined
by the above equation has to be even in order for the system to possess
the full B\"{a}cklund symmetry.
One finds now the general Lax to be:
\br \lefteqn{
L^{(2N+1)}_n \!=\! \( \pa - \vo_{n-1} + \sum_{p=1}^{N} \vp_{2n-2p} \( \pa -
\vo_{n-2p-1} - \vo_{n-2p} \)^{-1}\cdots \( \pa -
\vo_{n-3} - \vo_{n-2} \)^{-1} \) }   \lab{biglax}\\
& &\!\!\!\!\!\!\!\!\!\! \( \pa - \vo_{n-1} - \vo_{n} \)
\( \pa - \vo_{n} + \sum_{p=1}^{N} \vp_{2n-2p+1} \( \pa -
\vo_{n-2p} - \vo_{n-2p+1} \)^{-1}\cdots \( \pa -
\vo_{n-2} - \vo_{n-1} \)^{-1} \)       \nonu
\er
It transforms according to
\be
L^{(2N+1)}_{n+1} = \O_n L^{(2N+1)}_n \O_n^{-1}
\lab{biglaxtrans}
\ee
where $\O_n$ is defined by $\l^{1/2}  \Phi_{n+1} =\O_n \Phi_{n} $
and can be found from \rf{genspec} to be:
\be
\O_n=\( \pa - \vo_{n} + \sum_{p=1}^{N} \vp_{2n-2p+1}
 \( \pa - \vo_{n-2p} - \vo_{n-2p+1} \)^{-1}\cdots
 \( \pa - \vo_{n-2} - \vo_{n-1} \)^{-1} \) \lab{omegadef}
\ee
For the special case of 6-bosons Toda system with $a_0,a_1, a_2,a_3$ we define
the spectral equation on the square-lattice as
\be
\l^{1/2} \; \Phi_n =  \Phi_{n+1} + \vo_n  \Phi_{n-1}+ \vu_{n-2}  \Phi_{n-3}
\lab{specsix}
\ee
and get $ \l \Psi_n =L^{(6)}_n \Psi_{n}$ with
\br
L^{(6)}_n \eq \( \pa - A^{(0)}_n + ( \pa - B^{(0)}_{n} - A^{(0)}_{n+1} )^{-1}
B^{(1)}_{n}  \)  \( \pa - B^{(0)}_{n} - A^{(0)}_{n} \)^{-1}
\nonu\\
&& \!\!\!\( \pa - B^{(0)}_{n}+ A^{(1)}_{n} ( \pa - B^{(0)}_{n-1} -
A^{(0)}_{n}  )^{-1}\) \lab{sixlax}
\er
This Lax transforms according to
$$
{\wti L}^{(6)}_n = \( \pa - B^{(0)}_n+ A^{(1)}_{n} ( \pa - B^{(0)}_{n-1} -
A^{(0)}_{n} )^{-1}\) L^{(6)}_n  \( \pa - B^{(0)}_n+ A^{(1)}_{n}
( \pa - B^{(0)}_{n-1} - A^{(0)}_{n} )^{-1}\)^{-1}
$$
One notices checking corresponding equations of motion
that the Lax in \rf{sixlax} contains 6 independent
modes: $A^{(0)}_n, B^{(0)}_n, A^{(1)}_n, B^{(1)}_n$ together with
$A^{(0)}_{n+1}, B^{(0)}_{n-1}$.
By setting e.g. $B^{(1)}_n=0$ we would be left with four independent modes:
$A^{(0)}_n, B^{(0)}_n, A^{(1)}_n$ together with $A^{(0)}_{n+1}$.
This would describe four-boson Lax in terms of Volterra-like variables
but this time $g$-symmetry would be absent as the original lattice model has
an odd number of site functions ($a_0,a_1,a_2$).
\lskip
{\large {\bf 5. Comments}}
\lskip
Let us briefly comment on few remaining problems which we presently find
worth further investigations.

Recall, that in case of the two-boson system there was an interesting relation
between the second bracket structure of the system associated with the Toda
lattice and the one associated with the square-lattice. It turned namely out
that the gauge transformation from the Toda to the Volterra system served as
a Darboux transformation abelianizing the second bracket structure  of the
continuous Toda system.
It would be very helpful for the future investigation of the higher
bracket structures if this tendency would continue for the multi-boson system.
It would provide namely a convenient tool for deriving the second bracket
structure for all multi-boson hierarchies.

Our results show existence of the additional B\"{a}cklund
transformation for systems defined on the square-lattice and generalizing
Volterra systems to arbitrary even number of bosons.
By construction the $g$-symmetry can be associated only to original Toda
systems having even number of sites functions $a_0,a_1, \ldots, a_{2k+1}$.
Interestingly for the odd number of $a$'s the square-lattice system can
still be defined with Lax which is gauge equivalent to the original Toda
type Lax.
It would be interesting to investigate how the remaining features of the
generalized Volterra systems (like the bracket structures)
are affected by the apparent different symmetry content present in these
models.

Summarizing, we have seen how the continuum multi-boson KP hierarchy defines
through the gauge/similarity symmetry the Toda lattice model.
The more obvious transition from the Toda lattice to the continuum
multi-boson KP hierarchy leads from the Toda lattice equations to
representations of $\Win1$ via first bracket structure of the multi-boson KP
hierarchy.
One can ask a question whether the lattice versions of $\Win1$ can be obtained
by discretization via the gauge/similarity symmetry of the continuum model.

\lskip
{\bf Acknowledgements}
We thank E. Nissimov and S. Pacheva for comments on the manuscript
and J.F. Gomes for discussions.

\end{document}